\documentclass[12pt,preprint]{aastex}

\usepackage[]{emulateapj5}

\slugcomment{To be published in ApJ, v565, Jan 2002}

\shorttitle{Recent star formation in the halo of NGC~5128}
\shortauthors{Rejkuba et al.}

\begin{document}

\title{The radio/optical alignment and the recent star formation 
associated with ionized filaments in the halo of NGC~5128 (Centaurus A)
\thanks{Based on observations collected at the European Southern 
    Observatory, Paranal, Chile, within the Observing Programme 63.N-0229,
    and on observations collected by Magellan~I telescope 
    at Las Campanas Observatory, Chile.}}

\author{M. Rejkuba\altaffilmark{2}}
\affil{European Southern Observatory, Karl-Schwartzschild-Strasse
           2, D-85748 Garching, Germany}
\email{mrejkuba@eso.org}

\author{D. Minniti and F. Courbin}
\affil{Department of Astronomy, P. Universidad Cat\'olica, Casilla
        306, Santiago 22, Chile}
\email{dante@astro.puc.cl, fcourbin@astro.puc.cl}

\and

\author{D.R. Silva}
\affil{European Southern Observatory, Karl-Schwartzschild-Strasse
           2, D-85748 Garching, Germany}
\email{dsilva@eso.org}

\altaffiltext{2}{On leave from Department of Astronomy, P. Universidad Cat\'olica}

\begin{abstract}
We used the direct CCD camera at the 
Magellan~I telescope at Las Campanas Observatory
and FORS1 at Antu VLT at ESO Paranal Observatory to
image fields centered on the inner and outer optical filaments
in the halo of NGC~5128. In the $V$ vs. $U-V$ color-magnitude diagrams we  have 
identified young blue supergiants associated with these 
line-emitting filaments located between the inner radio
lobe and the northern middle lobe. Around the outer filament stars as young 
as 10~Myr were detected. They are principally aligned with the direction 
of the radio jet, 
but a vertical north-east alignment along the edge of the H~{\sc i} 
cloud is also present. Young stars in the inner filament field are found inside 
the bright knots of photoionized gas and are strongly aligned in the 
direction towards the center of the galaxy at the same position angle as
the inner radio jet. Fitting the Padova isochrones on UV color-magnitude 
diagrams we find that the blue stars around the inner filaments have ages 
similar to the ones around the outer filaments $\sim10-15$~Myr and the same 
abundance of  Z=0.004. The presence of young blue supergiants clearly 
shows that the bright blue knots in the north-eastern halo 
of NGC~5128 are associations of young stars with photoionized gas. 
The temperature of the brightest stars is T$\sim 12000-16000$~K, 
insufficient to account alone for the high excitation lines observed 
in the surrounding ionized gas. Thus the optical emission jet is principally
seen due to its alignment with the radio structure of the AGN. 
The highly collimated star formation is present only in the north
eastern halo of the galaxy, suggesting the interaction of the jet with 
the gas clouds deposited during the last accretion event as the 
preferred triggering mechanism. From these observations we infer a lower
limit for the age of the NGC~5128 jet of $10^7$~yr. 
The triggering of the star formation in
the dense clouds in the halo of the galaxy by the jet supports the
alignment effect observed in high redshift radio galaxies. It also suggests
that radio galaxies should have higher than normal star formation rates.  
\end{abstract}


%
%-------------------------------------------------------------------
%

\section{Introduction}

NGC~5128, the nearest giant elliptical galaxy \citep{israel}, was associated
with the strong radio source Centaurus~A by \citet{bolton49}. In the optical,
the galaxy has an elliptical light distribution crossed in the center by a
strong warped dust lane. The diffuse optical extensions were first
discovered by \citet{johnson63} and subsequently the rich system of shells
around the elliptical galaxy was detected by \citet{malin83}. 
\citet{blanco75} and \citet{peterson75} discovered the optical 
jet or filaments in NGC~5128. The origin and the nature of 
the emission-line filaments have been hotly debated
\citep{dufour,gp81,graham83,morganti+91,sutherland+93,morganti+99}.

The origin of diffuse shells \citep{malin83} can be explained with phase
wrapping of stars from an accreted disk galaxy \citep{quinn84}. 
\citet{rejkuba+01} discussed the stellar populations in
the prominent diffuse north-eastern shell and in the halo of NGC~5128. The
`outer filament' situated at $13\farcm6$ from the centre extends for
$\sim 8\arcmin$ in the direction of position angle $22^\circ$
\citep{blanco75} overlaping partially with the diffuse stellar shell. While
the stellar populations in the shell are similar to the rest of the halo of
the giant elliptical, the stars close to the emission-line filament are
predominantly young \citep[see also][]{mould00,fg00}. 

The `inner filament' is located at the distance of $7\farcm8$ from the
nucleus and extends outward over $\sim2\arcmin$ at a position angle
$55^\circ$ \citep{blanco75}. It consists of three bright knots, identified as 
A, B and C by \citet{blanco75} and two more diffuse structures, E and F 
\citep{morganti+91}. 
Here we report the detection of a young stellar population
associated with the optical emission-line inner ($r\sim8\arcmin$) 
and outer ($r\sim14\arcmin$) filaments in the north-eastern halo 
of NGC~5128. Throughout the paper we adopt distance modulus to 
NGC~5128 of 27.8 \citep{soria+96,hhp99} that 
corresponds to a distance $D=3.6$~Mpc. At this distance, $1\arcmin$ 
on the sky translates to 1.047 kpc in the galaxy.

%
%--------------------------------------------------------------------
%

\section {The Data}
\label{data}

We used FORS1 at VLT Antu to image the outer field 
because it is less crowded and the filaments cover a larger area 
than in the inner field.
On the contrary in the inner field Magellan~I was necessary because despite 
of having a smaller area it gives higher resolution for the crowded field.

The field with the inner filaments ($\alpha_{2000}=13^h26^m05.9^s$,
$\delta_{2000}=-42^{\circ}56\arcmin35\arcsec$)
was imaged with the Magellan~I (Walter Baade) Telescope, one 
of the twin 6.5 m Magellan telescopes at Cerro Manqui at the Las Campanas
Observatory. The observations were taken on 22 February 2001, 
during the first Chilean run on the new telescope and one of the first 10 
nights of scientific operations. 
The instrument was the direct CCD camera Tektronix $2048 \times 2048$ 
with gain of 2 $e^{-}/DN$. 
At the f/11 focus of Magellan~I the pixel scale is
$0\farcs069$/pixel with a total field size of $2\farcm36^2$. 
The small pixel scale and the  extraordinary good and stable 
seeing of $0\farcs31-0\farcs38$ during the 
whole night enabled us to obtain well sampled point spread function (PSF) 
images with almost no crowding problems deep in the halo of NGC~5128. 
The observations consist of $3\times 1800 {\rm s}$, 
$2\times 900 + 30 {\rm s}$ and $2\times300 {\rm s} + 4\times900 
{\rm s}$ exposures in Harris $U$-, $V$- and $I$-bands\footnote{The exact
names of filters were LC-3012, LC-3014 and LC-3011. For the
filter transmission curves see
$http://www.lco.cl/lco/instruments/manuals/direct\_ccd/filters/$.}, 
respectively.
The color composite image of the field can be found at the following URL
$http://www.lco.cl/magellan\_lco/publicity/science\_images/cena.html$.

The standard stars from the \citet{landolt} catalogue 
in two different fields were observed at four different 
airmasses on 22 and 23 February 2001 with the identical 
instrumental setup. Additional photometric standards in these fields 
were defined by \citet{stetson00} in the $V$ and $I$ bands, 
making the total number of useful 
standard stars observed 15, 50 and 48 in $U$, $V$ and $I$, respectively. The 
following calibration equations reproduce the calibrated magnitudes with 
one-sigma scatter around the mean of 
0.08, 0.024 and 0.031 for $U$, $V$ and the $I$ band, respectively. 
\begin{eqnarray}  
u_{\rm inst} & = & U - 24.771(\pm0.168 ) + 0.36 (\pm0.15 )*X \nonumber \\
         &   & -0.064(\pm0.011)*(U-V)
\label{Utran}
\end{eqnarray}
\begin{eqnarray}
v_{\rm inst} & = & V - 26.756(\pm 0.005)  + 0.15 (\pm 0.05)*X \nonumber \\
         &   & + 0.045 (\pm 0.004)*(V-I)
\label{Vtran}   
\end{eqnarray}
\begin{eqnarray}  
i_{\rm inst} & = & I - 26.390(\pm0.008 ) + 0.06 (\pm0.03 )*X \nonumber \\
         &   & -0.068(\pm0.007)*(V-I)
\label{Itran}
\end{eqnarray}
 
The observations of the field that contains 
outer filaments ($\alpha_{2000}=13^h26^m23.5^s$,
$\delta_{2000}=-42^{\circ}52\arcmin00\arcsec$) 
were carried out in the 
service mode with FORS1 instrument on the 8.2m ESO Very Large 
Telescope (VLT) Antu at Paranal, on July 11$^{th}$ and 12$^{th}$ 1999. They 
consist of pairs of 15-min optical (Bessel $U$- and $V$-band) 
exposures taken in photometric conditions with seeing ranging from
$0\farcs45$ to $0\farcs55$. The FORS1 detector has pixel scale of
$0\farcs2$/pixel and the field of view $6\farcm8\times6\farcm8$. 
A detailed description of these data and their 
reduction is given in \citet{rejkuba+01}. 

The photometric measurements of the outer filament images were described in 
\citet[][Field~1]{rejkuba+01}. The photometry 
in the inner filament field followed a similar procedure. The stars were 
identified on the median combined image with the 
FIND algorithm that is part of the 
DAOPHOT~II programme  \citep{stetson}. The PSF magnitudes were 
measured on all individual frames simultaneously with ALLFRAME \citep{stetsonALF}. 
Restricting the PSF fitting parameters sharpness 
and chi we rejected bad pixels and extended objects like galaxies and 
globular clusters \citep{rejkuba01} from the final photometry list.
Only the stars with good photometry ($\sigma_{mag}< 0.5$) in at least two images
were kept. The final photometric catalogue of of the inner filament field contains 
1612, 8003 and 7683 stars with good photometry in the $U$, $V$ and $I$ bands, 
respectively.
%
%----------------------------------------------------------------------
%

\section{The Color-Magnitude and Color-Color Diagrams}
\label{CMD}

\begin{figure}
\plotone{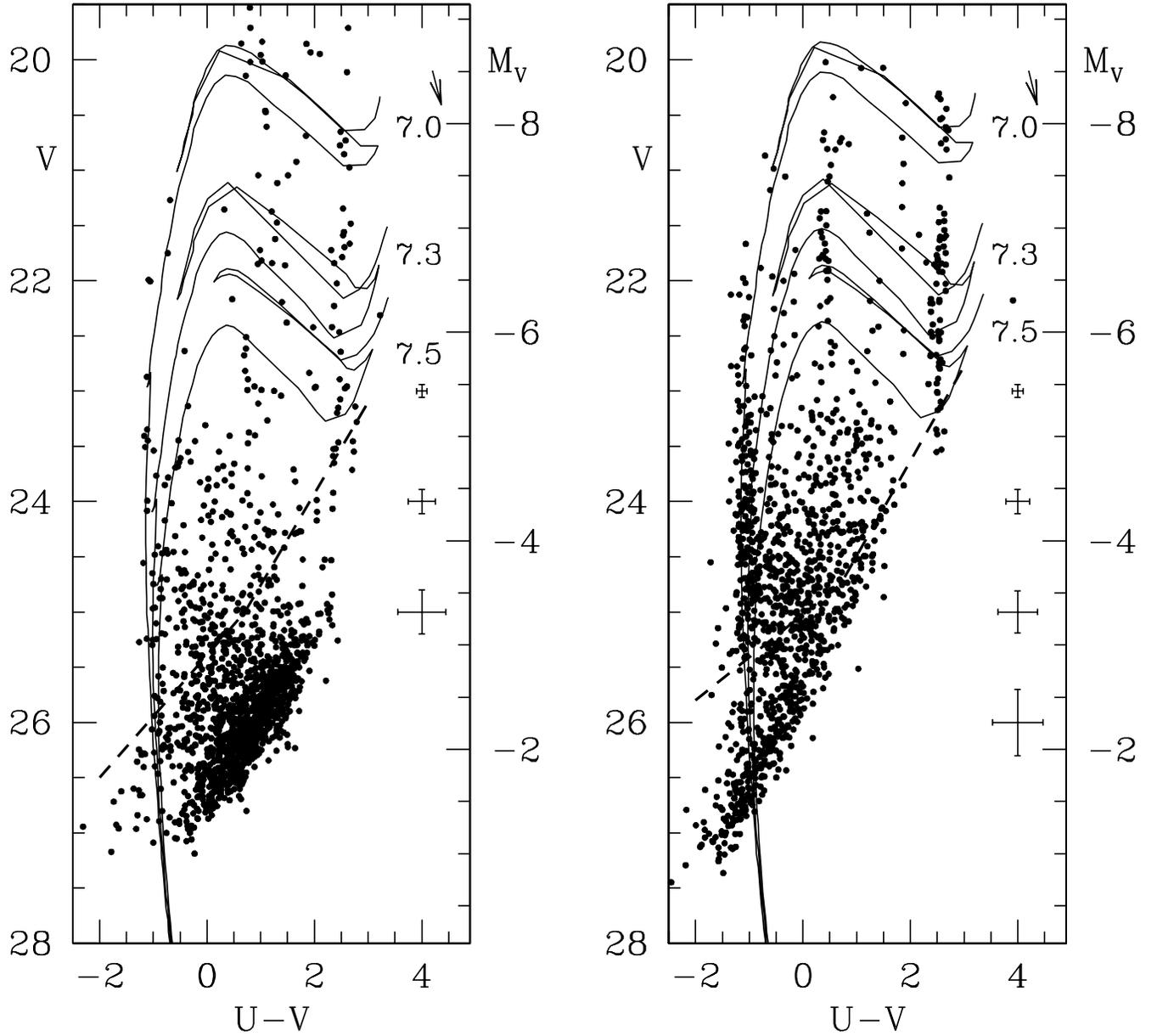}   
  \caption[]{UV color-magnitude diagram for the inner filaments field (left) 
and the outer filament field (right). The foreground stellar contamination 
was statistically subtracted.
The arrow represents the reddening vector of E(B$-$V)=0.1. Typical
photometry errors are plotted on the right side. The dashed line indicates 50\% 
completeness limit. Full lines are isochrones from \citet{bertelli94} for 
Z=0.004. The 
logarithm of ages in years is indicated on the right edge of each isochrone.}
  \label{UVcmds}
\end{figure}

In Fig.~\ref{UVcmds} we compare the UV color-magnitude diagrams of the inner and
outer filament fields. The much larger field of VLT than Magellan~I 
and the more extended star formation region in the outer field accounts for 
higher number of blue stars with colors $(U-V)\sim-1$. In both fields there 
is a young blue main sequence extending to magnitudes of $M_V\sim-7$. 
We ``cleaned'' the CMDs from foreground contamination using the 
Besan\c{c}on group model of stellar populations in the Galaxy 
\citep{robin&creze,robin}. Some of the stars in the red 
part ($U-V>0$) of the CMD are most probably still the 
remaining Galactic foreground stars, but in that area of the 
CMD supergiants on their blue--red excursions are also located (see also
\citet{rejkuba+01}). 

The width of the blue to red 
supergiant loops in the red part constrains the abundance, while the brightness 
of the main sequence stars constrains the age of the population. 
In order to determine an upper limit for the temperature and ionizing 
fluxes of the stars we assumed the age and abundance from the isochrone fitting. 
The best fit from the available set of Padova isochrones \citep{bertelli94} 
is obtained for Z=0.004 and the log(age$)\ge 7.0$~yr.  The brightest stars 
in both fields are lying along the $10^7$ yr old isochrone. Assuming then 
that the brightest stars have Z=0.004 and log(age)=7.0-7.1 yr, the
temperature of the most luminous stars is between $\sim12000$ and $16000$~K.  

The comparison with another halo field in the southern region of the galaxy 
(Field~2 in \citet{rejkuba+01}) indicates that the 
young population is confined to the regions aligned with the optical filaments. 

\begin{figure}
\plotone{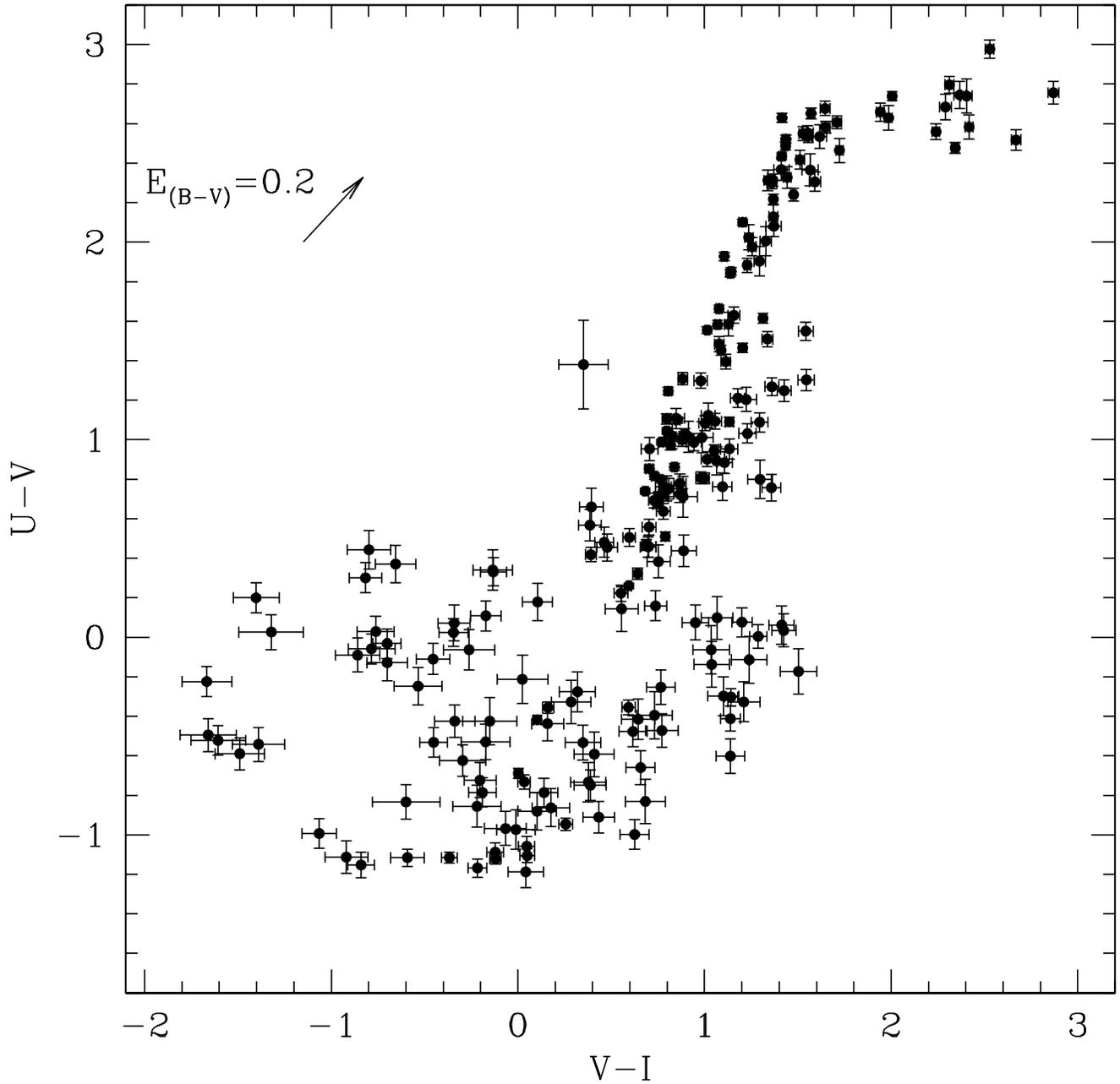}   
  \caption[]{Color-color diagram for the point sources matched in U, V and
$I$ frames of the inner filament field 
that had ALLFRAME photometry errors $<0.5$ mag and magnitudes
brighter than 50\% completeness limits in all bands.
}
  \label{ccdUVI}
\end{figure}

The (U-V) vs. (V-I) color-color diagram for all the stars in the 
inner filament field with photometric errors $<0.5$ mag and magnitudes 
brighter than 50\% 
completeness limits in all bands is presented in Fig.~\ref{ccdUVI}. 
The foreground stars have not been subtracted. They form the sequence that
crosses the diagram from upper right towards lower left corner. 
The objects in the lower left part of the color-color diagram are 
young supergiants in NGC~5128. The spread is due to possible differential
reddening and spread in ages of the young stars. Most galaxies have $0<(U-V)<4$ 
and $0.25<(V-I)<1.25$ \citep{rodighiero}, so only few of them at 
higher redshift may 
contribute to the spread in the bluest part of the color-color diagram. 
The isolated point to the left of the stellar sequence at $(U-V)=1.4$ and
$(V-I)=0.4$ is most probably a galaxy. Absence of other points with similar
colors indicate a good selection criteria and low contamination by background 
galaxies in our stellar photometry.

The reddening value inferred from the isochrone fitting to the 
stellar sequence is 
$E(B-V)=0.10\pm0.05$  in good agreement with the values from the literature 
\citep{schlegel,fg00,rejkuba+01}. The width of the young blue sequence is 
somewhat larger in the inner than in the outer filament field. This 
may indicate the presence of the small differential reddening 
amounting at most to 
$E(B-V)=0.07$. However, a possible spread in ages of young stars and 
the photometric errors at fainter magnitudes complicate the interpretation.

%                                                                      
%----------------------------------------------------------------------
%

\section{Knot A - Image deconvolution}

\begin{figure}
\plottwo{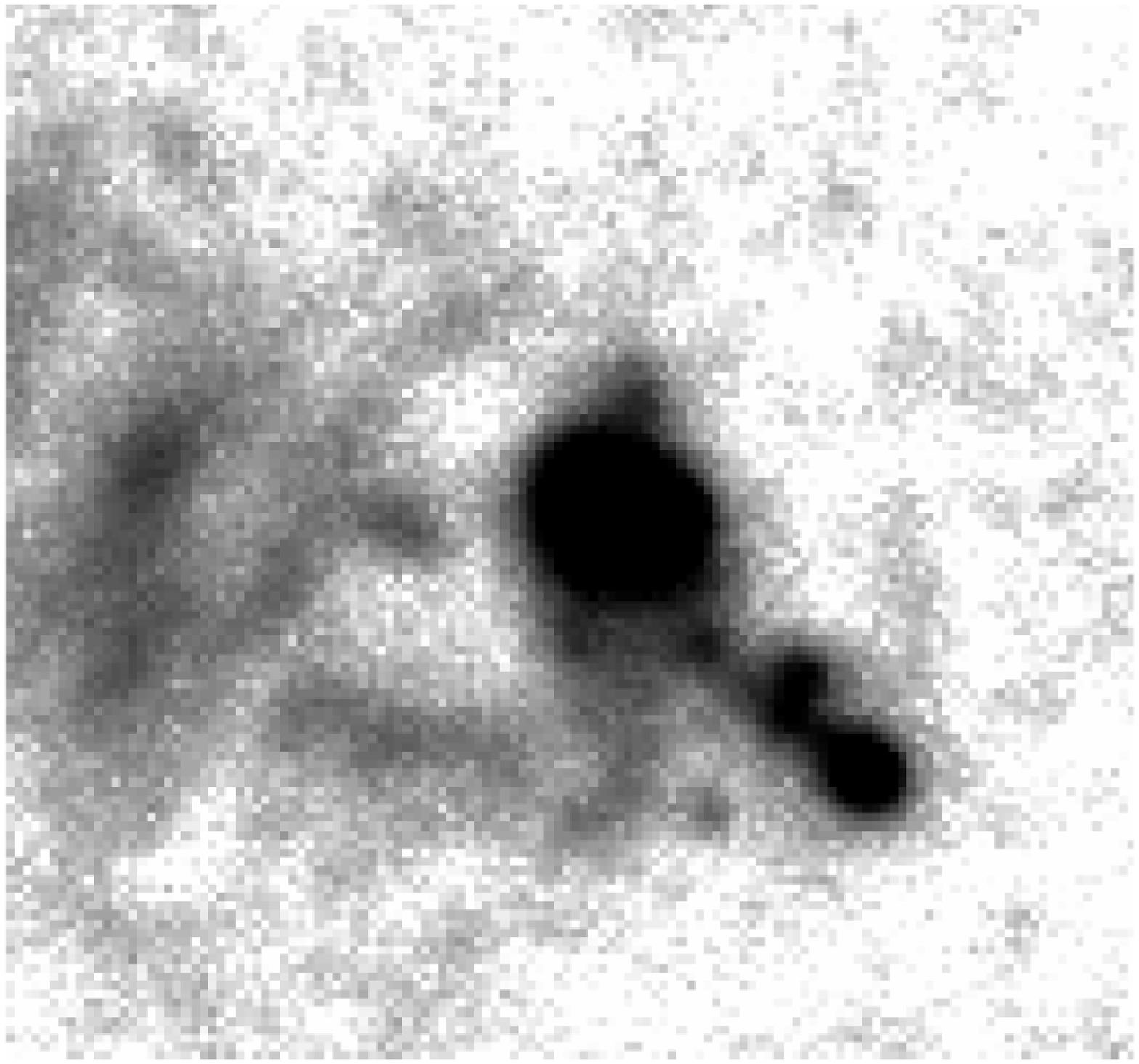}{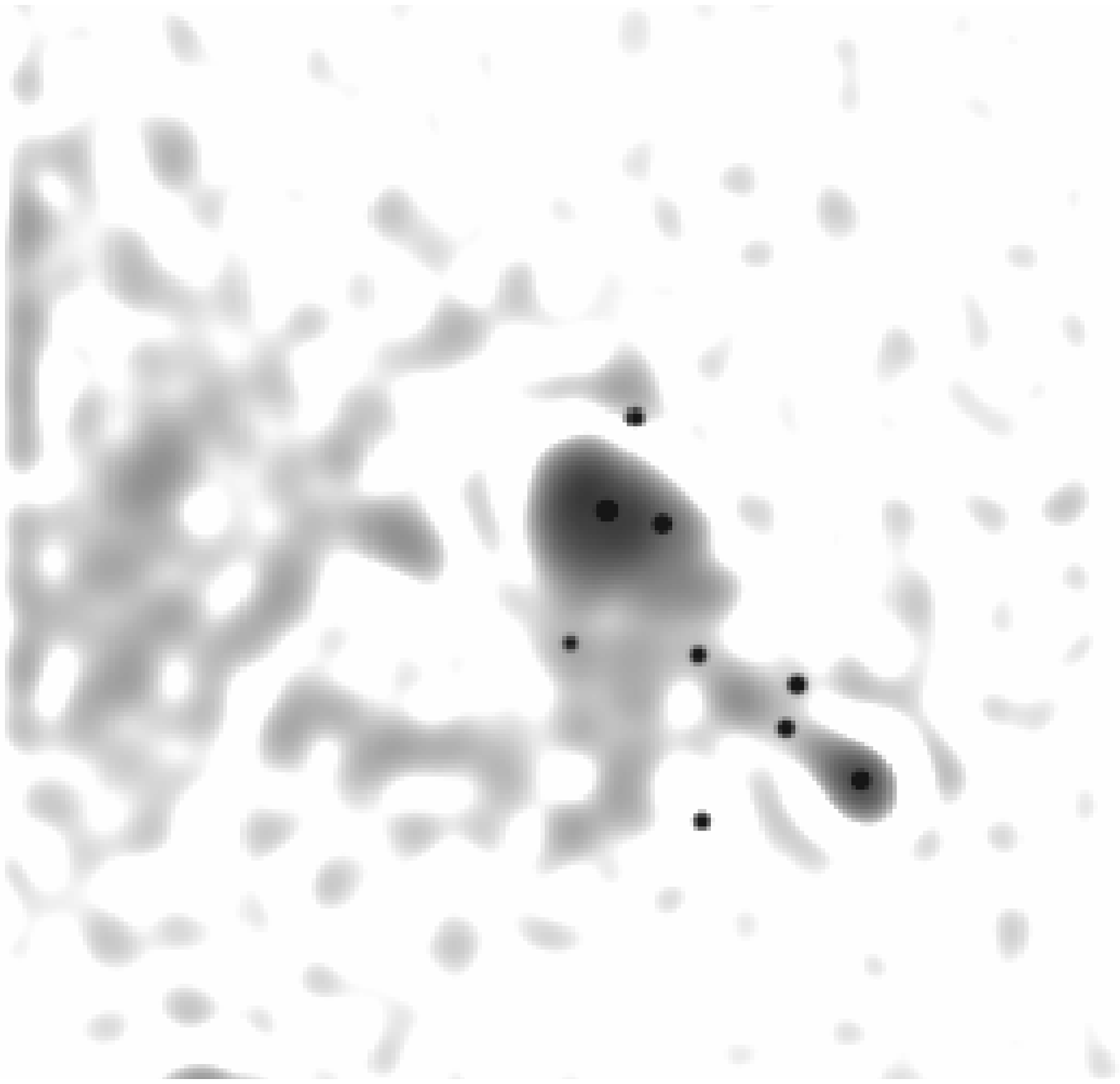} \\
\plotone{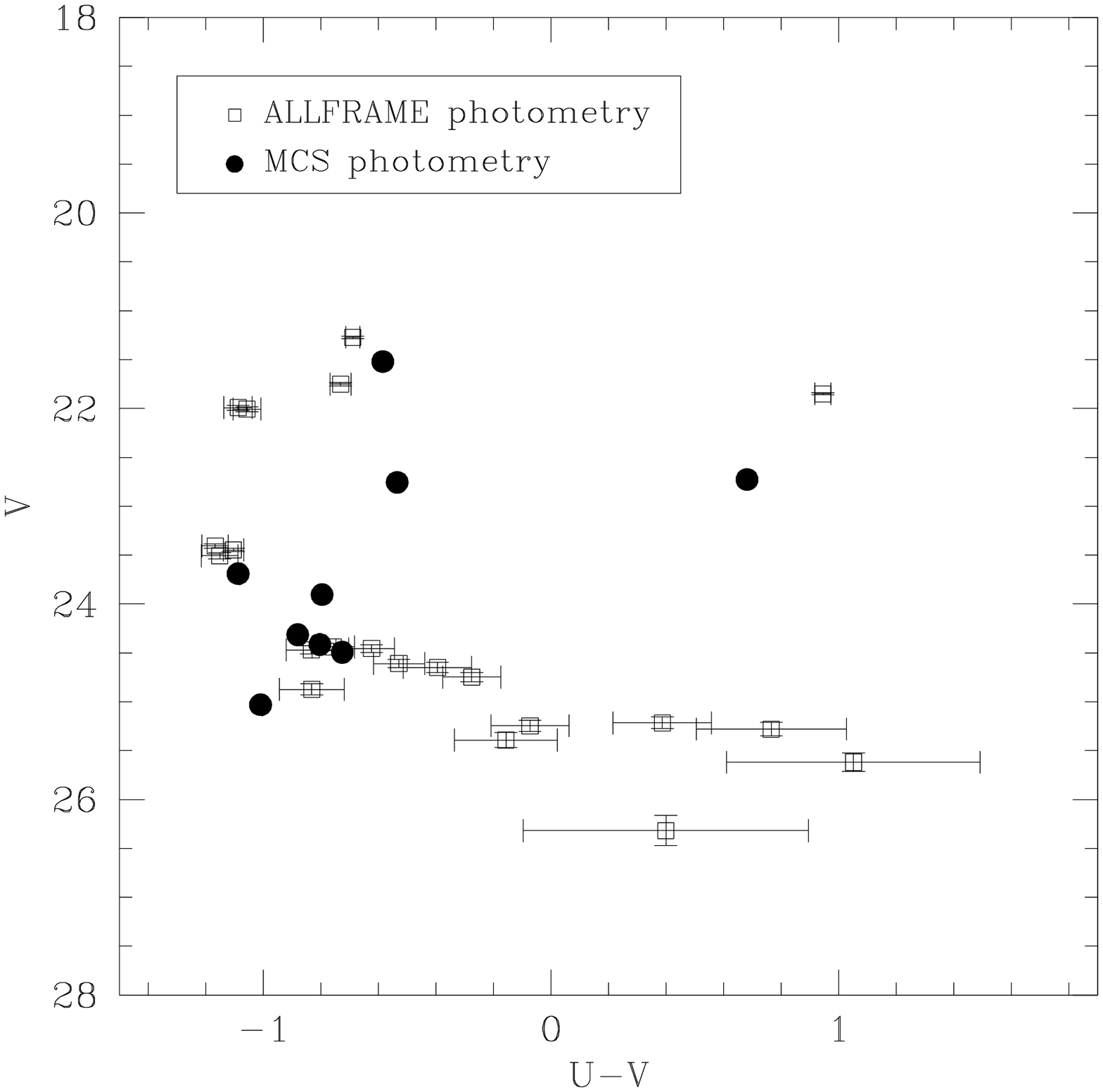}
  \caption[]{Top left: A detail of the inner filament field 
centered on knot A. This is the combined $V$-band stamp of 
$9\arcsec \times 9\arcsec$. North is up and east to the left.
Top right: The same image as on the left 
deconvolved with the MCS algorithm. In the CMD under the 
images the photometry obtained with ALLFRAME from the original frames is compared 
with the photometry measured on deconvolved images.}
  \label{knotA}
\end{figure}

The photometry of knot A \citep{blanco75}, 
a possible compact star forming region in the jet,
required particular care. It consists of a mixture of stellar like objects, 
extended nebular emission and an unresolved background of faint stars. 
In order to estimate the relative contribution of these components to 
our photometry, we deconvolved a small portion of the Magellan~I image 
of $9\times9$ arcsec around the brightest hot spot using the
MCS deconvolution algorithm \citep{MCS}. In this way we obtained a new
image with improved spatial resolution (Fig.~\ref{knotA}, top right) 
and a background map (or the ``extended source channel'') 
containing the extended sources. 
The final pixel scale on the deconvolved image is  
$0\farcs0345$ and the resolution of point sources is $0\farcs069$.

In the MCS algorithm, one has to give as an input, initial estimated 
positions and intensities for all suspected point sources. 
The program then iteratively changes them while trying to
deconvolve the data. The difference between the data and the 
deconvolved image (reconvolved back with the PSF) in units of the
noise, is used to check the quality of the process, a posteriori. 
This difference image or residual map should be flat and equal to 1 
everywhere in the field in case of perfect deconvolution.
Since introducing a higher number of point sources in the image will always 
produce a better result, we deliberately choose the minimum number of point
sources required for the deconvolution process to yield acceptable residual 
maps. Some faint stars in the deconvolved image are therefore present in the 
``extended source channel'' of the data and we ``see'' fewer stars with the 
deconvolution process than with ALLFRAME. However, the stars we see are 
decontaminated from the extended underlying nebular emission and
we can infer their magnitudes and colors with no contribution from 
the emission of the surrounding gas. 
The comparison between these new measurements 
and the photometry obtained with ALLFRAME  
(Fig.~\ref{knotA}) indicates that the magnitudes and colors of 
blue stars measured with ALLFRAME are not excessively contaminated by 
line emission from the surrounding gas.
In particular we show that knot A is 
composed of at least two point sources embedded in a more
extended envelope, possible nebular emission or a very compact and still
unresolved cluster. Other objects in the field do not show such extended 
emission apart from the compact object $\sim3$ arcsec SW of knot A.
The residual flux of knot A after star subtraction is 
$9.4 \times 10^{-17}$ erg/cm$^2$/sec/\AA.

%                                                                      
%----------------------------------------------------------------------
%
\section{Discussion}
\label{discussion}

Many high-redshift radio galaxies show optical images elongated in 
the same direction as their double radio sources, the so called alignment 
effect \citep{mccarthy+87,chambers+87,debreuck}. The alignment has often been 
explained with models of star formation induced by shocks associated with 
the passage of radio components \citep{rees,daly,bestetal}, but in some
cases scattering of nuclear light by dust or electrons and 
line-emission excited by the continuum from the obscured active nucleus 
are dominating optical and UV light \citep{diserego+89,dickson+95,vernet+00}.
In NGC~5128 we can observe in more detail 
physical processes that are acting in these more distant galaxies.

\begin{figure}
\plotone{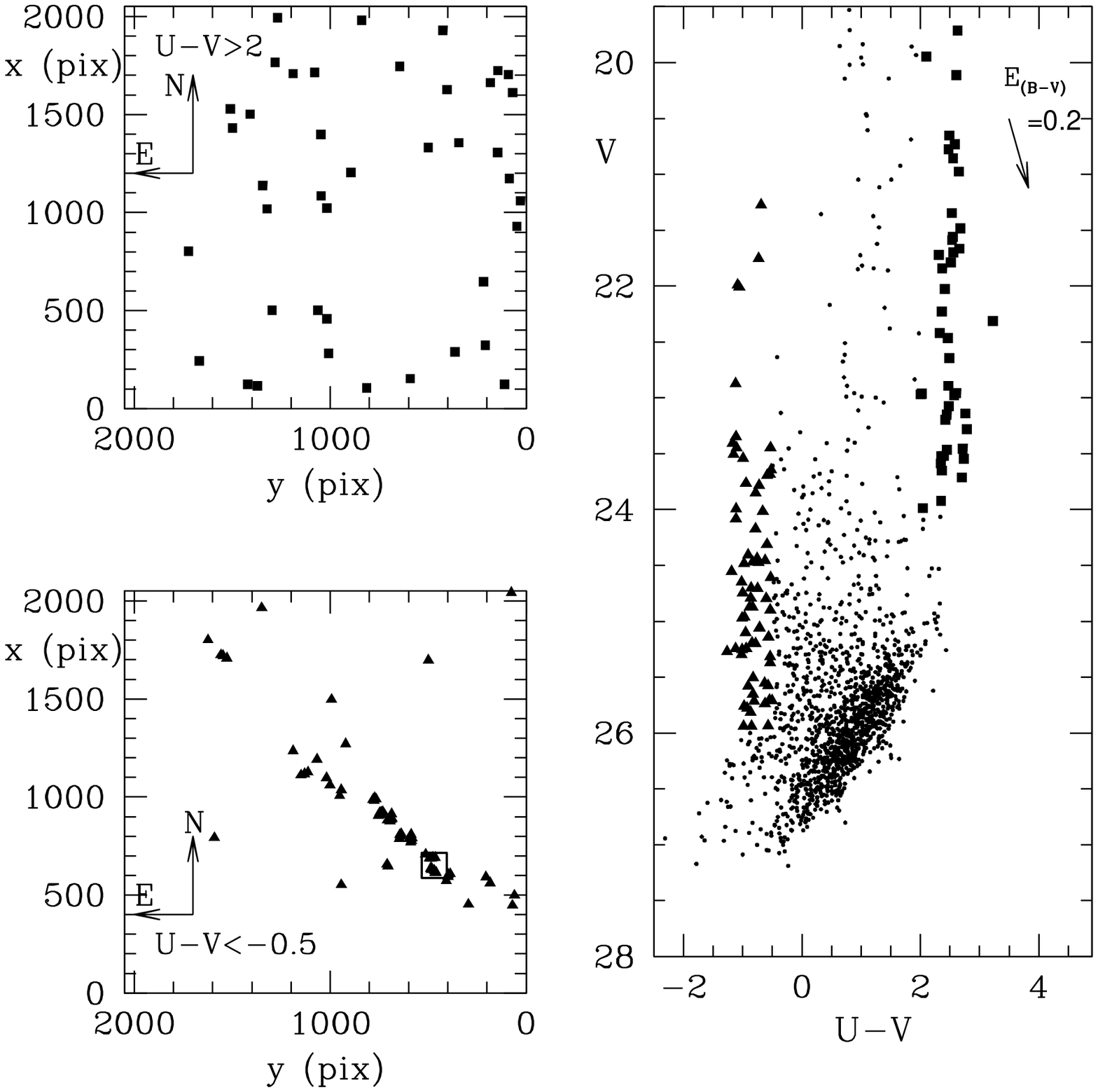}
  \caption[]{The spatial distribution of the bluest (U$-$V)$<-0.75$ (triangles)
and the reddest (U$-$V)$>2$ (squares) stars in the inner filament field  
 for stars with magnitudes above the 50\%
completeness limit. The field of view is $2\farcm4\times2\farcm4$ and the 
scale $0\farcs069$/pixel. 
The box indicates the location of knot A. 
The positions of the stars in the respective UV CMD
are indicated on the right.}     
  \label{blue_red_inner}
\end{figure}

\begin{figure}
\plotone{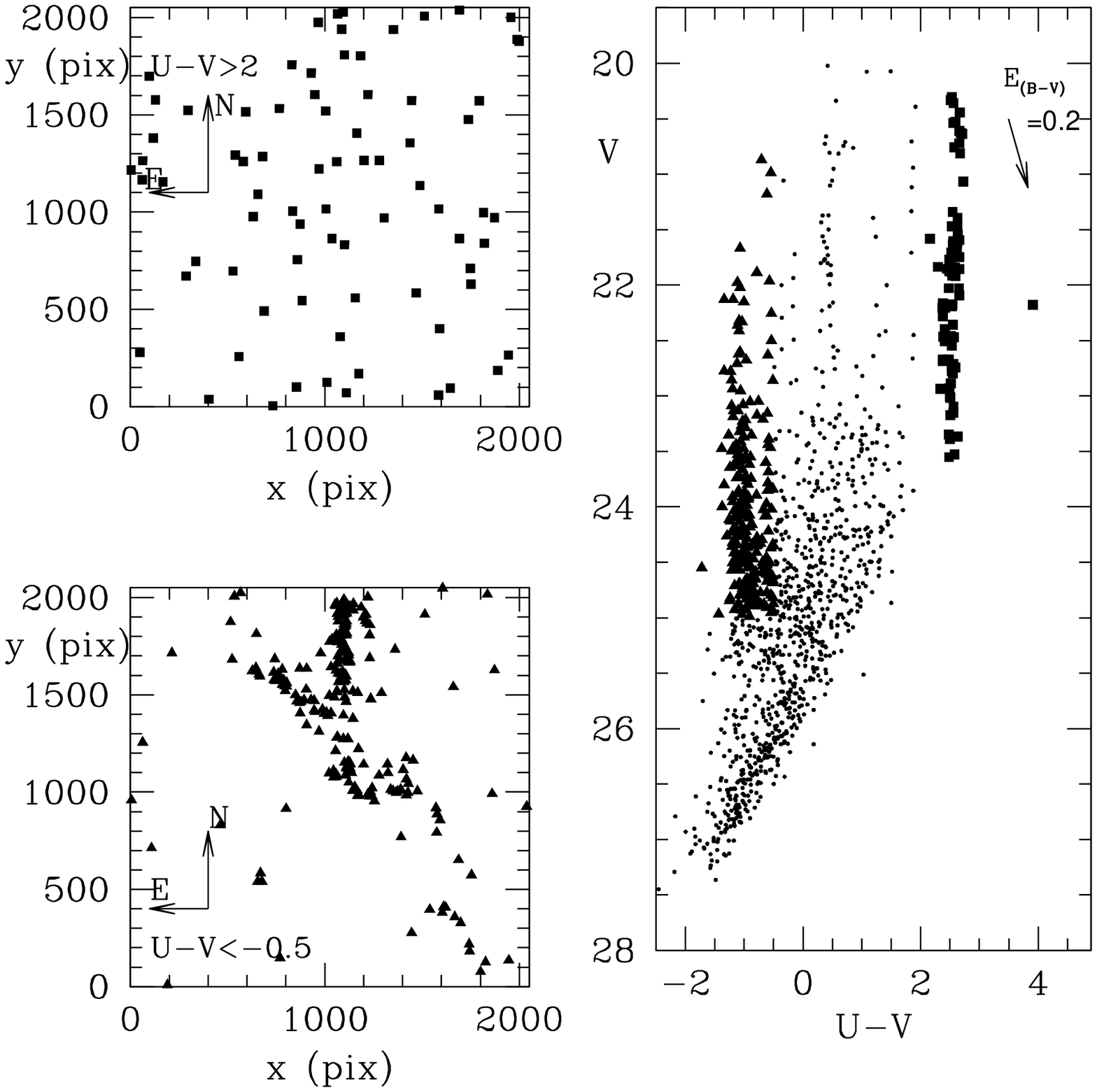}
 \caption[]{The spatial distribution of the bluest (U$-$V)$<-0.75$
(triangles)
and the reddest (U$-$V)$>2$ (squares) stars in the outer filament field for
stars with magnitudes above the 50\%
completeness limit. The field of view is $6\farcm8\times6\farcm8$ and the 
scale $0\farcs2$/pixel. 
The positions of the stars in the respective UV CMD
are indicated on the right.}
  \label{blue_red_outer}
\end{figure}

The direct observations of young blue supergiants associated with the outer 
filaments were recently reported by several authors 
\citep{fg00,mould00,rejkuba+01}. Now for the first time we have 
resolved the young stars associated with the ionized filaments and 
knots in the inner part of the galaxy, and mapped these young stars over 
a large field along the radio jet. The remarkable alignment of 
the young stars and ionized gas  over $>10$~kpc 
is shown in Fig.~\ref{blue_red_inner} and \ref{blue_red_outer}. 
The red stars ($U-V>2$) are evenly distributed in both fields as 
expected for foreground galactic stars, while almost all recently 
formed stars lie along the knots and filaments of ionized gas. 
The north-south (vertical) alignment in the outer field is along the 
eastern edge of the H~{\sc i} cloud that is associated with the stellar 
shells \citep{schiminovich}. 

We checked the locations of all of the blue stars and in particular 
of the most luminous ones. In the outer field they are evenly 
distributed along the star forming region, while the four brightest blue 
stars in the inner field lie within a very small region, 
known as knot A \citep{blanco75}. \citet{gp81} 
found that the spectrum of the knot A shows characteristics of 
a normal H~{\sc ii} region and conclude that there must be hot 
stars embedded in that gas. 
We resolved here the brightest stars in that H~{\sc ii} region. 
The northern extension
of knot A is distinguished by a higher excitation and electron 
temperature similar to what is observed in 
other knots and filaments in the inner and outer halo 
\citep{gp81,morganti+91}. 
The high excitation lines observed to be prominent in the filaments 
in the halo of NGC~5128 (e.g. He~{\sc ii} $\lambda$4686 and high ratio of 
[O{\sc iii}] $\lambda$5007/[O{\sc ii}] $\lambda$3727$>1$) 
show that the stars cannot be the only source of ionizing photons. 

The complex velocity field of the filaments creates a puzzle. 
The spectroscopy of the ionized gas revealed a very wide range of 
velocities covering more than 550 km s$^{-1}$ 
\citep{gp81,graham83,morganti+91,graham98} and varying significantly on
scales of $\la 100$~pc. The mean velocity of the gas, around $\sim 300$ 
km s$^{-1}$, is a few hundred km s$^{-1}$ 
lower than the systemic velocity of the central region 
\citep{graham79,morganti+91,graham98}, but is close to the measured velocity 
of the H~{\sc I} cloud that is located to the NW of the outer filaments 
\citep{schiminovich}. If the stars shared the turbulent 
velocities of the gas, in $10^7$ yr they would travel $\sim 3$~kpc, 
which correspond to approx. $3\arcmin$ at the distance of NGC~5128 
(note that the field of view of Magellan is 
less than $2\farcm5^2$ and that of VLT $\sim 6\farcm7^2$). 
If they had these velocities, the gas and
the stars would have been dispersed in the halo of the galaxy. 

The large amount of the H~{\sc i} and CO gas 
\citep{schiminovich,charmandaris} and its association with 
the diffuse stellar shells in the halo can be interpreted 
with the dynamical scenario of phase-wrapping, following the merger of a 
spiral galaxy with NGC~5128 approximately $10^8$ yr ago 
\citep{quinn84,charmandaris}. 
The similar mean velocity of the H~{\sc i} cloud and the ionized gas in the 
filaments indicates the common origin of the gas. We detect here
for the first time the young stars associated with the inner
filaments. Their strong alignment with the line emitting filaments
and with the radio and x-ray jet coming from the AGN nucleus of NGC~5128 
suggests a connection between them and poses interesting questions
like ``Did the jet trigger the star formation?'' or 
``Are the jets seen because of the star formation?'' and 
``Would we see the jet if there were no blue stars there?''. 

The answer to the first question is more complex and thus we address
the others first. If we assume that the brightest stars have ages 
of $\sim10$~Myr and metallicities of Z=0.004, then they have 
temperatures of $\sim 15000$~K. The high excitation lines
observed in the spectra of the ionized gas require ionization 
temperatures of the order of $10^5$~K. It is clear that an additional 
ionization mechanism is exciting the gas. The alignment of the ionized 
gas with the direction towards the nucleus of 
the galaxy and the fact that even though it is present, the gas in the other 
parts of the halo of NGC~5128 is not ionized, leaves as the most simple 
possibility either an interaction with the radio jet that leads to the 
shock excitation \citep{sutherland+93} or photoionization by the 
radiation field of a nuclear continuum source 
\citep{morganti+91,morganti+92}. In any case, the ``jet'' of 
optical emission line features is not seen due to the star formation 
and it would most probably be visible even if the stars were not there. 

The young stars formed after the collapse of the gas probably have the 
same origin as the H~{\sc I} and molecular clouds \citep{charmandaris}. The 
presence of this cool gas, with $T\la10^4$~K, is the necessary ingredient
in the jet induced star formation models. In the model described 
by \citet{rees} the prime effect of the radio outburst 
in a galaxy would be to heat the surrounding medium and evacuate the radio
lobe of thermal gas. However, it would not disturb the cool-phase clouds, which 
would then be triggered to collapse in the overpressured medium. 
So, the jet could have induced the collapse and the star 
formation without transferring to the stars the high turbulent velocities 
that are observed today in the gas. The star formation process driven by the
external pressure may not be dissimilar from the one observed in the
classical stellar associations, such as Orion.
This means that one needs to find an 
alternative mechanism in order to explain the velocities of the gas 
that have been observed. 
One possibility is that supernova explosions induced shocks into
the surrounding medium. However, the problem with this scenario is 
that the first generation of supernovae would explode some $10^6$ yr 
after the onset of the star formation. So, the gas would acquire 
the high velocities still $\sim8-9$  million years ago and would 
have had plenty of time to disperse. 
It would have too low density and would probably not be visible any more. 
Thus either the first supernovae were not efficient enough to blow away 
all the gas or some new gas came from elsewhere. 
Also, the jet that triggered the star formation would have induced high
velocities in the hot gas, which would then have plenty of time to disperse 
in the halo of the galaxy. 
The obvious source of the new gas is the close-by  H~{\sc I} cloud 
that will continue to supply fresh material that can be ionized 
as long as it stays in the direction of the jet. 

The inner and the outer filaments are not mutually aligned. Similarly 
the inner and the outer radio jets are misaligned. If we assume that 
the jet is precessing as suggested also by \citet{hce83}, it should 
have first hit the outer filament and precess in the eastward direction
hitting later the inner filament field. The recent observations of the 
radio jet with VLBI \citep{tingay+98} and the 
x-ray jet with Chandra \citep{kraft+00} show close alignment 
with the inner filaments at the position angle of $55^\circ$. 
Stars in the inner and the outer filament have the same ages to within our 
precision, of the order of $\sim 1-2 \times 10^6$~yr. 
The distance between the inner and the outer filaments is $\sim20^\circ$, 
implying the projected precession rate of the the order of $10^{-5}$ 
degrees per year. The presence of the large quantity of neutral gas 
towards the west of the outer filament implies that either the jet is precessing
in a cone whose right edge does not pass the outer filaments on the west or 
that it is intermittent and started shining only ${\rm few} \times 10^7$~yr ago. 

Given the facts that the blue stars are aligned along the direction of 
the jet, that they have formed recently, that there is a lot of gas all 
over the halo of the galaxy \citep{schiminovich}, but new stars are formed 
only where the jet passes across or close to that gas, we can conclude 
that the young stars observed in the filaments in the NE halo of 
NGC~5128 are result of the jet induced star formation. 
The optical emission from the gaseous filaments associated with 
those stars is mainly powered by the radiation field of the hidden nuclear
continuum source \citep{morganti+91,morganti+92} or by the shock excitation
due to the interaction of the radio jet with a dense cloud of material at the
location of the filaments \citep{sutherland+93}.
If this cause-effect scenario is correct, we can put a lower limit to the 
age of the jet itself: it must have been acting for at least $10^7$~yr 
in order to form the stars observed.

%                                                                      
%----------------------------------------------------------------------
%
\section{Conclusions}
\label{conclusions}

We have used the direct camera at the Magellan~I and FORS1 at Antu VLT to
measure the stellar photometry in two fields in the halo of the giant elliptical
galaxy NGC~5128. The fields were located at the position of the inner and outer
optical filaments \citep{blanco75}. We detected the chain of blue stars
aligned with the line-emitting filaments and with the radio jet. The main
conclusions from this work are:

\begin{itemize}

\item[1.] We see young blue stars aligned with the jet over $>10$~kpc. 

\item[2.] The young stars have ages of $10^7$~yr and their UV
color-magnitude diagram is best fit with 
Padova isochrones \citep{bertelli94} for abundance of Z=0.004.

\item[3.] The jet triggered star formation in the halo of NGC~5128, presumably
when hitting dense clouds of neutral gas left in the halo  by the previous 
accretion event \citep{schiminovich,charmandaris}. 

\item[4.] The main excitation mechanism that ionizes the gas associated with
the young stars is either beamed photoionization from the obscured 
AGN \citep{morganti+91,morganti+92} or shock ionization by the radio jet 
\citep{sutherland+93}. The young stars with temperatures of the order of
$\sim 15000$~K contribute only a smaller part of ionizing photons.

\item[5.] Our results support the alignment effect observed in 
high redshift radio galaxies, and suggest that radio galaxies should 
have higher than normal star formation rates.

\item[6.] There is evidence for the jet to be processing from the outer
to inner filament field eastward in the sky.

\item[7.] We infer a lower limit for the age of the jet as $10^7$ yr.

\end{itemize}

\acknowledgments
We are grateful to R. Fosbury for stimulating discussion and valuable 
comments that improved this paper and to P. Stetson for the careful reading of 
the manuscript. We thank the staff of Las Campanas 
Observatory for their help during the observations as well as the ESO 
astronomers who took the VLT observations for us in service mode. 
MR acknowledges ESO studentship programme.
This work was performed in part under the auspices of the Chilean Fondecyt
No.\ 01990440.

\clearpage

\clearpage

\clearpage

\clearpage

\end{document}